\documentclass[twoside,12pt]{article}

\def\ds{\displaystyle}

\begin{document}
\renewcommand{\refname}{\begin{center}{\normalsize\rm REFERENCES}\end{center}}
\begin{center}
\large{NUMERICAL TREATMENT OF HYPERSINGULAR INTEGRAL EQUATIONS
WITH APPLICATION TO CRACK MECHANICS}
\end{center}

\markboth{\hfill Iovane G., Sumbatyan M.A.\hfill } {\hfill
Hypersingular integral equations with applications \hfill}

\addcontentsline{toc}{subsection} {Iovane G., Sumbatyan M.A.
Numerical Treatment of Hypersingular Integral Equations with
Application to Crack Mechanics}

\begin{center}
\large{\bf Iovane G.$^{*}$, Sumbatyan M.A.$^{**}$}

 \normalsize{\it $^{*}$ University of Salerno, Italy} \\
 \normalsize{\it $^{**}$ Rostov State University, Russia }
\end{center}

\vspace{\baselineskip} \small In this paper we present a treatment
of hypersingular integrals and integral equations. It is shown the
equivalence of different definitions of hypersingular integrals.
Then we construct explicit analytical solution to a characteristic
equation and prove convergence of the collocation method to this
solution. Further, we expand these results to full hypersingular
kernels, and give a demonstration of the developed method to some
problems of crack mechanics including crack problem for porous
elastic materials, in the framework of Cowin-Nunziato model.

\vspace{\baselineskip}
\normalsize

{\bf 1.~The method.} ~ Let us consider the hypersingular equation
of the following type
$$
\int\limits_{-1}^{1}\varphi \left( t\right) \left[ \frac{1}
{\left(x-t\right) ^{2}}%
+K_{0}\left( x,t\right) \right] dt=f\left( x\right) ,\qquad \qquad
\left| x\right| <1~, \eqno(1.1)
$$
where\ $\ K_0(x,t)$ \ is a regular part of the kernel. Direct
numerical treatment of Eq.(1.1) is not easy. A particular case is
the hypersingular equation with the characteristic kernel
$$
\int\limits_a^b\frac{g\left( t\right) dt}{\left( x-t\right)
^{2}}=f' \left( x\right) ,\qquad \qquad x\in \left( a,b\right)
,\quad f\left( x\right) \in C_{2}\left( a,b\right)~. \eqno(1.2)
$$
We find a solution of this equation bounded on the both ends
$x=a,b$. The latter can be defined in exact explicit form, which
is given by the following inversion formula
$$
g\left( x\right) =\frac{\sqrt{\left( x-a\right) \left( b-x\right) }}{\pi ^{2}%
}\int\limits_{a}^{b}\frac{f\left( t\right) dt}{\sqrt{\left(
t-a\right) \left( b-t\right) }\,\left( x-t\right) }~. \eqno(1.3)
$$
This result shows that  any bounded solution of Eq.(1.2) vanishes at $%
x\rightarrow a,b.$ To construct a direct collocation technique to
numerically solve equation (1.2) for arbitrary right-hand side, we
divide the interval $\left( a,b\right) $ to $n$ small equal
subintervals by the nodes
$a=t_{0},\,t_{1},\,t_{2},...,t_{n-1,\,}t_{n}=b.$ Then
the length of each interval is $h=\left( b-a\right) /n,$ and the nodes $%
t_{j}=a+jh,\quad j=0,1...,n.$ Let us denote the central points of
each sub-interval $\left( t_{i-1},t_{i}\right) $ by $x_{i}$, so that $%
x_{i}=a+\left( i-1/2\right) h\,,~ i=1,...,n.$ Therefore, we try to
approximate Eq.(1.2) by the linear algebraic system
$$
\sum_{j=1}^{n}g\left( t_{j}\right) \left( \frac{1}{x_{i}-t_{j}}-\frac{1}{%
x_{i}-t_{j-1}}\right) =f^{\prime }(x_i),\qquad \qquad i=1,...,n.
\eqno(1.4)
$$
It is proved in [1] that by assuming $x\in \left( a,b\right) $,
the difference between solution $g\left( x\right) $ of the system
(1.4) at the point $x$ and solution given by explicit formula
(1.3) tends to zero, when $n\rightarrow \infty $. It can be proved
that system (1.4) admits the explicit solution
$$
\begin{array}{c}
\ds g\left( x_{l}\right) =\frac{\Delta _{l}}{\Delta }=\left(
x_{e}-t_{0}\right) ~\sum_{m=1}^{n}~
\frac{\sum\limits_{k=1}^mf^{\prime }\left( x_{k}\right) }
{t_{m}-t_{0}}\times  \vspace*{4mm}\\
\ds\times \frac{\prod\limits_{p}\left( x_{l}-t_{p}\right) ~
\prod\limits_{q}\left( x_{q}-t_{m}\right) ~} {\left(
x_{l}-t_{m}\right) ~ \prod\limits_{p\neq m}\left(
t_{m}-t_{p}\right) ~ \prod\limits_{q\neq l}\left(
x_{q}-x_{l}\right) }~.
\end{array}
\eqno(1.5)
$$
If the number of nodes increases then $g(x_{l}) \rightarrow g(x)$.

Let us we consider the full equation
$$
\int\limits_{a}^{b}
\left[ \frac{1}{\left( x-t\right) ^{2}}+K_{0}\left( x,t\right) %
\right] g\left( t\right) dt=f^{\prime }\left( x\right) ,\quad x\in
\left( a,b\right)~ , \eqno(1.6)
$$
where
$$
K_{0}\left( x,t\right) =\frac{\partial K_{1}\left( x,t\right)
}{\partial x}. \eqno(1.7)
$$
Its bounded solution can be constructed by applying inversion of
the characteristic part, that reduces Eq.(1.6) to a second-kind
Fredholm integral equation:
$$
g\left( x\right) +\int\limits_{a}^{b}N_{1}\left( x,t\right)
g\left( t\right) dt=f_{1}\left( x\right) ,\quad x\in \left(
a,b\right)~ , \eqno(1.8)
$$
where
$$
N_{1}\left( x,t\right) =\frac{\sqrt{\left( x-a\right) \left( b-x\right) }}{%
\pi ^{2}}\int\limits_{a}^{b}\frac{K_{1}\left( \tau ,t\right) d\tau
} {\sqrt{\left(\tau -a\right) \left( b-\tau \right) }~ \left(
x-\tau \right) }~, \eqno(1.9)
$$
$$
f_{1}\left( x\right) =\frac{\sqrt{\left( x-a\right) \left(
b-x\right) }}{\pi ^{2}}\int\limits_{a}^{b} \frac{f\left( \tau
\right) d\tau}{\sqrt{\left( \tau -a\right) \left( b-\tau \right)
}~\left( x-\tau \right) }~. \eqno(1.10)
$$

It is known from the classical theory of the Cauchy-type integrals
that if $f\left( x\right) \in C_{1}\left( a,b\right) ,$
$K_{1}\left( x,t\right) \in C_{1}\left[ \left( a,b\right) \times
\left( a,b\right) \right] $ then also $f_{1}\left( x\right) \in
C_{1}\left( a,b\right);~ N_{1}\left( x,t\right) \in C_{1}\left[
\left( a,b\right) \times \left( a,b\right) \right] $. Further, we
prove that if $f\left( x\right) \in C_{1}$ $(a,b);~K_{1}\left(
x,t\right) \in C_{1}\left[ \left( a,b\right) \times \left(
a,b\right) \right] $, then for any $x\in \left( a,b\right) $ the
difference between solution $g\left( x\right) $ of the linear
algebraic system
$$
\sum\limits_{j=1}^{n}\left[ \frac{1}{x_{i}-t_{j}}-
\frac{1}{x_{i}-t_{j-1}}+ hK_0\left( x_{i},t_{j}\right) \right]\,
g\left( t_{j}\right) =f^{\prime }\left( x_{i}\right) ,~i=1,...,n
\eqno(1.11)
$$
and the bounded solution of equation (1.6) tends to zero when
$h\rightarrow 0 $ (i.e. $n\rightarrow \infty $).

{\bf 2.~Application to crack mechanics.} ~ The linear
Cowin-Nunziato theory of homogeneous and isotropic elastic
material with voids is described by the following system of
partial differential equations (where $\phi =\nu -\nu _{0}$ is the
change in volume fraction from the reference one [2,3])
$$
\left\{
\begin{array}{c}
\mu ~\Delta ~\bar{u}+(\lambda +\mu )~{\rm grad~div}~\bar{u}+\beta ~{\rm grad}%
~\phi =0\vspace*{3mm} \\
\alpha ~\Delta ~\phi -\xi ~\phi -\beta ~{\rm div}~\bar{u}=0\quad ,
\end{array}
\right. \eqno(2.1)
$$
where $\mu $ and $\lambda $ are classical elastic constants;
$\alpha ,~\beta
$ and $\xi $ -- some constants related to porosity of the medium. Besides, $%
\bar{u}$ denotes the displacement vector.
The components of the stress tensor are defined, in terms of the functions $%
\bar{u}$ and $\phi $, by the following relations ($\delta _{ij}$
is the Kronecker's delta)
$$
\left\{
\begin{array}{l}
\sigma _{ij}=\lambda ~\delta _{ij}~\varepsilon _{kk}+2~\mu
~\varepsilon
_{ij}+\beta ~\phi ~\delta _{ij} \\
\varepsilon _{ij}=\displaystyle\frac{1}{2}(u_{i,j}+u_{j,i})~.
\end{array}
\right. \eqno(2.2)
$$

If we consider a plane-strain boundary value problem for the thin
crack of the length $2a$ with plane faces, dislocated over the
segment $-a<x<a$ along the $x$-axis. Let the plane-strain
deformation of this crack be caused by a normal load $%
-\sigma_0 $ symmetrically applied to the faces of the crack. For
the last problem the boundary conditions over the line $y=0$ are
$$
\sigma_{xy}=0~,~\frac{\partial\phi}{\partial
y}=0~(|x|<\infty)~,\quad \sigma_{yy}=-\sigma_0~(|x|<a)~,\quad
u_y=0~(|x|>a)~, \eqno(2.3)
$$
where
$$
c^{2}=\frac{\mu }{\lambda +2\mu }~,\quad H=\frac{\beta }{\lambda +2\mu }%
~,\quad l_{1}^{2}=\frac{\alpha }{\beta }~,\quad
l_{2}^{2}=\frac{\alpha }{\xi }~,\eqno(2.4)
$$
with the first two numbers $c~,~H$ being dimensionless and the quantities $%
l_{1}~,~l_{2}$ -- of dimension of length.

Let us apply the Fourier transform along the x-axis to relations
(2.1)--(2.3). Then the problem can be reduced to the following
integral equation:
$$
\int\limits_{-b}^{b}g(\xi )K(x-\xi )d\xi =-\,(1-N)^{2}\,\frac{\sigma _{0}}{%
2\mu }~,\qquad |x|<b~.\eqno(2.5)
$$
where $g$ is the opening of the crack face, $N=(l_{2}^{2}/l_{1}^{2})\,H~,%
\quad (0\leq N<1)$ and
$$
\begin{array}{c}
\displaystyle K(x)=\frac{1}{2\pi }\int\limits_{-\infty }^{\infty }\frac{|s|}{%
q(s)}[2Nc^{2}s^{2}(q-|s|)+(1-N)(1-N-c^{2})q]e^{-isx}ds\vspace*{4mm} \\
\displaystyle=\frac{1}{\pi }\int\limits_{0}^{\infty }L(s)\cos
(sx)ds,\qquad
q=q(s)=\sqrt{s^{2}+1-N}~,\vspace*{4mm} \\
\displaystyle L(s)=\frac{s}{q(s)}\left[ 2Nc^{2}s^{2}(q-s)+(1-N)(1-N-c^{2})q%
\right] ~.
\end{array}
\eqno(2.6)
$$
The kernel (2.6) admits explicit representation by special
functions that permits direct estimate of its singular properties.
It can be seen that the kernel is hypersingular, and we apply the
proposed method to numerically solve equation (2.5). Then we study
in detail the influence of the porosity of the material to the
stress concentration coefficient.

\vspace{\baselineskip}

{\bf Acknowledgment}\par\noindent This work was supported in part
by the G.N.F.M. of the Italian Research Council (C.N.R. Italy) and
by a Russian Leading School in Mechanics (Grant 2113.2003.1).


\begin{thebibliography}{99}
\small
\bibitem{1}  {\it G.\,Iovane, I.K.\,Lifanov, M.A. Sumbatyan \/} On direct numerical
treatment of hypersingular integral equations arising in mechanics
and acoustics // Acta Mechanica (accepted).
\bibitem{2}  {\it S.C.\,Cowin, J.W.\,Nunziato \/} Linear elastic materials with
voids // J. Elasticity. 1983. V.~13. P.~ 125--147.
\bibitem{3}  {\it A.\,Scalia, M.A.\,Sumbatyan\/} Contact problem for porous
elastic half-plane // J. Elasticity. 2000. V.~60. P.~91--102.

\end{thebibliography}
\end{document}